\documentclass[twocolumn,showpacs,preprintnumbers,amsmath,amssymb]{revtex4}

\usepackage{graphicx}
\usepackage{dcolumn}
\usepackage{bm}

\begin{document}

\title{Temporary cooling of quasiparticles and delay in voltage response of superconducting bridges after abrupt switching on the supercritical current}

\author{D.Yu. Vodolazov$^{1,2}$}
\email{vodolazov@ipmras.ru}
\author{F.M. Peeters$^3$,}
\affiliation{$^1$ Institute for Physics of Microstructure,
Russian Academy of Sciences, 603950,
Nizhny Novgorod, GSP-105, Russia \\
$^2$ Lobachevsky State University of Nizhny Novgorod,
23 Gagarin Avenue, 603950 Nizhny Novgorod, Russia \\
$^3$Departement Fysica, Universiteit Antwerpen (CGB),
Groenenborgerlaan 171, B-2020 Antwerpen, Belgium}

\date{\today}


\begin{abstract}

We revisit the problem of the dynamic response of a superconducting bridge after abruptly switching on
the supercritical current $I>I_c$. In contrast to previous theoretical works we take into account spatial gradients and use both the
local temperature approach and the kinetic equation for the distribution function of quasiparticles. In both models the finite delay time
$t_d$ in the voltage response is connected with temporary cooling of quasiparticles due to the suppression of the superconducitng order
parameter by current. We find that $t_d$ has different values and different temperature dependencies in the considered models.
In turns out that the presence of even small inhomogeneities in the bridge or of bulk leads/contacts at the ends of the {\it homogenous}
bridge favors a local suppression of the superconducting order parameter $|\Delta|$ during the dynamic response. It results
in a decrease of the delay time, in comparison with the spatially uniform model, due to the diffusion of nonequilibrium quasiparticles from
the region with locally suppressed $|\Delta|$. In case the current distribution is spatially nonuniform across the bridge the delay time
is mainly connected with the time needed for the nucleation of the first vortex at the position where the current density is maximal
(at $I\sim I_c$ and for not very wide films). We also find that a short alternating current pulse (sinusoid like) with zero time-average may result
in a nonzero time-averaged voltage response where its sign depends on the phase of the ac current.

\end{abstract}

\maketitle

\section{Introduction}

In 1979 Pals and Wolter \cite{Pals} observed a long delay (about hundreds of nanoseconds) in the appearance of the voltage response after the instant
(on a time scale of 1 ns) application of the supercritical current to an Al superconducting film. This work initiated a large number of studies
(both experimental and theoretical) which aimed to clarify the physical origin of this effect
\cite{Frank,Geier,Wolter,Attekum,Wolter2,Tinkham_review,Jelila}.
The main conclusion was that the supercritical current suppresses the superconducting order parameter $\Delta=|\Delta|exp(i\phi)$ and
its decay in time provides some kind of temporary 'cooling' of quasiparticles \cite{Tinkham_book,Tinkham_review}. Because the critical current
in superconductors increases, when the temperature decreases, this effective cooling shifts the applied current $I>I_c(T)$ closer to the
nonequilibrium $I_c$ (corresponding to lower 'temperature')
and it slows down the destruction of superconductivity. It was experimentally observed that the time delay $t_d$ decreases fast with increasing
amplitude of the current and in some experiments a strong dependence of $t_d$ on temperature was found \cite{Frank,Wolter,Attekum,Wolter2}
while in other works $t_d$ practically did not depend on T \cite{Pals,Jelila}.

In the majority of  previous theoretical studies on this subject authors assumed that superconductivity decays
uniformly in space \cite{Pals,Geier,Tinkham_review} which considerably simplified the analytical calculations. Besides it was assumed
that the nonequilibrium quasiparticle distribution function $f(\epsilon)$ is not thermal (i.e. it cannot be described by the Fermi-Dirac
distribution function with an effective temperature and chemical potential) and that the phonons are in
equilibrium. These assumptions become invalid when the inelastic electron-electron relaxation time $\tau_{e-e}$ is of the same order
as the inelastic electron-phonon relaxation time $\tau_{e-ph}$ and/or if the escape time of the nonequilibrium phonons $\tau_{esc}$ to
the substrate is larger than $\tau_{e-ph}$. In the present work we extend the model of Tinkham \cite{Tinkham_review}
(which we correct quantitatively by taking a proper expression for the energy derivative of the equilibrium $f(\epsilon)$) to the spatially
nonuniform case and to the case when the deviation from equilibrium could be described in terms of a local temperature (quasi-equilibrium limit
\cite{Nagaev,Giazotto}). The last model predicts the temperature independent delay time, which was earlier
observed in some experiments \cite{Pals,Jelila}, and which was not explained by previous theoretical works.

In addition we answer the question: how the time delay is modified when the current density distribution is nonuniform in the superconductor.
Nonuniformity may come from the Meissner (screening) effect which is important in wide films/bridges with width $w>\lambda^2/d$ ($\lambda$ is
the London penetration depth, $d$ is the thickness of the film/bridge). In narrow films/bridges ($w\ll \lambda^2/d$) spatially nonuniform current
distribution may arise due to the current crowding effect near edge/surface irregularities \cite{Buzdin,Aladyshkin,Vodolazov_SB} or due to
specific geometry \cite{Clem_crowding}. Below we show, that for not very wide bridges and currents close
to $I_c$, the delay in the voltage response is mainly connected with the appearance of the first vortex and qualitatively resembles the
time delay of quasi 1D bridges. By application of weak magnetic field one may tune the current distribution in the film and change the time delay.
We also discuss the relation of this problem with the recent activity devoted to superconducting single photon detectors (SSPD) \cite{SSPD_review}.

We also investigate the response of the superconducting bridge on a short alternating current pulse. Our interest to this problem arises
from a recent experiment where the voltage response of a wide YBCO superconducting bridge to a short pulse of synchrotron radiation
was investigated \cite{Probst}. As proposed in Ref. \cite{Probst} the short electromagnetic pulse (with duration of about several picoseconds)
induces a current pulse in the superconducting bridge. It was found that such a pulse does not substantially heat the superconductor but it
leads to a finite voltage response. This experiment demonstrates the possibility to study the resistive response of the superconducting bridge
on a very short time scale ($\sim$ several picoseconds) which is hard to realize by different methods.
In the framework of the used theoretical models we predict that a nonzero time averaged voltage response can be obtained from a zero time averaged
ac current pulse which we explain by the long delay time in the destruction of superconductivity by current.

The structure of our paper is as follows. In section II we present our theoretical model that we use to study the dynamic
response of 'dirty' superconductors near its critical temperature $T_c$. In section III we present the results of our calculations
for quasi-1D bridges in the non-thermal
(subsection IIIA) and quasi-equilibrium (subsection IIIB) cases. Subsequently, in section IV we present results for wide ($w\gg \xi(T)$)
superconducting bridges placed in a weak magnetic field (which creates a nonuniform current distribution) and in section V we study the dynamic
response of the superconducting bridge on a sinusoidal-like current pulse. In section VI we discuss the applicability of our results to
different physical situations.

\section{Model}

To model the dynamical response of the 'dirty' superconductor near $T_c$ we use the simplified set of equations which was derived
in Refs. \cite{Schmid1,Larkin,Kramer1,Watts-Tobin,Schmid2}. Near $T_c$ one may neglect the coupling between odd $f_L(\epsilon)$ and even $f_T(\epsilon)$
energy parts of the quasiparticle distribution function $f(\epsilon)=(1-f_L-f_T)/2$ and instead of solving the kinetic equation for $f_T(\epsilon)$
we use the simplest possible approximation (below we discuss the validity of this approach)
\begin{equation}
f_T=-e\varphi \frac{\partial f_L^0}{\partial \epsilon}
\end{equation}
where $f_L^0=tanh(\epsilon/2k_BT)$ and we assume a small deviation from equilibrium $\delta f_L=f_L-f_L^0\ll f_L^0$.

With these simplifications the equations for $f_L$ and $\Delta$ have the following form
\begin{equation}
N_1\frac{\partial f_L }{\partial t }=D\nabla((N_1^2-R_2^2)\nabla f_L)-\frac{N_1}{\tau_{in}}(f_L-f_L^0)-
R_2\frac{\partial f_L^0}{\partial \epsilon}\frac{\partial |\Delta|}{\partial t},
\end{equation}
\begin{eqnarray}
\frac{\pi\hbar}{8k_BT_c} \left(\frac{\partial }{\partial
t}+2ie\varphi \right) \Delta= \nonumber
\\
\xi_{GL}^2\left( \nabla -i\frac{2e}{\hbar c}A\right)^2\Delta+\left(1-\frac{T}{T_c}+\Phi_1-\frac{|\Delta|^2}{\Delta_{GL}^2}\right)\Delta,
\end{eqnarray}
where $\xi_{GL}^2=\pi\hbar D/8k_BT_c$ and $\Delta_{GL}^2=8\pi^2(k_BT_c)^2/7\zeta(3)$ are the zero
temperature Ginzburg-Landau coherence length and the order parameter correspondingly, $A$ is the vector potential, $\varphi$ is the electrical potential
and $\Phi_1=\int_0^{\infty}R_2 \delta f_Ld\epsilon/|\Delta|$. From Eq. (3) it follows that for the uniform case and in equilibrium
$|\Delta|=\Delta_{eq}=\Delta_{GL}(1-T/T_c)^{1/2}$.

To find the solution of Eq. (2) one should use the Usadel equation for the normal $\alpha (\epsilon)=\cos\Theta=N_1(\epsilon)+iR_1(\epsilon)$
and anomalous $\beta_1 =\beta e^{i\phi}$, $\beta_2=\beta e^{-i\phi}$ ($\beta (\epsilon)=\sin\Theta=N_2(\epsilon)+iR_2(\epsilon)$) Green functions
\begin{equation}
\left(\left(2i\epsilon-\frac{\hbar}{\tau_{in}}\right)-\frac{D}{\hbar} q_s^2 \cos\Theta\right)\sin\Theta+2|\Delta|\cos\Theta=0,
\end{equation}
where $q_s=mv_s=(\nabla \phi-2eA/\hbar c)$ is the superfluid momentum. In Eq. (4) we skip the term with the spatial derivative. We checked out that
its presence weakly affects the time delay at $T\geq 0.9 T_c$. In contrast, we find that the term proportional to $q_s^2$ leads to a
considerable decrease of the time delay.

Within the same approximation the current density in the superconductor may be written as
\begin{eqnarray}
j=\frac{\sigma_n}{e}  \frac{\pi|\Delta|^2q_s}{4k_BT_c}+\frac{\sigma_n}{e}\int_0^{\infty}j_{\epsilon} \delta f_L d\epsilon
-\sigma_n\nabla \varphi=
\nonumber
\\
=j_s+\delta j_s+j_n,
\end{eqnarray}
where the first term on the right hand side (RHS) is the superconducting current density, $j_n=-\sigma_n \nabla \varphi$ ($\sigma_n=2e^2DN_0$) is the normal current
density ($N_0$ is the one spin density of states at the Fermi level). In Eq. (5) we keep the nonequilibrium contribution to the supercurrent
($\delta j_s$) due to $\delta f_L\neq 0$. In Ref. \cite{Geier} it was argued that the presence of this term increases $t_d$ when $I\gg I_c$.

Let us now discuss when Eq. (1) is correct. From Eq. (3) (which is similar to the standard
time-dependent Ginzburg-Landau equation except for the term $\Phi_1 \Delta$ in the RHS) and $div j=0$ (where $j$ is defined by Eq. (5))
it follows that the conversion of the normal current to the superconducting one (at the normal metal-superconductor boundary)
occurs on the scale $L_E\simeq \xi(T)$, while it is well known that near $T_c$ this length is $L_E\simeq D\tau_{in} k_BT/|\Delta|)\gg \xi(T$ \cite{Ivlev}.
In our problem the spatial gradient of $j_n$ appears only at the end of the transition period, when $|\Delta| \to 0$ somewhere in the bridge,
and we do not expect any influence of a different $L_E$ on the time delay. Besides, already at $T\simeq 0.9 T_c$ the majority of the normal current
is converted to the superconducting one on a length scale $\xi(T)$ \cite{Hsiang} which is connected with an increased contribution
of Andreev reflection to the conversion of the current when the superconducting gap increases with decreasing temperature.

When the escape time of phonons to the substrate is shorter than the inelastic electron-phonon relaxation time and at the same time
the electron-electron scattering is weaker than the electron-phonon one then the relaxation time in Eq. (2) $\tau_{in}=\tau_{e-ph}$ and the
quasiparticle distribution function is not a thermal one. In this case Eqs. (2-4) and the current continuity equation $div j=0$ are the equations
that govern the dynamic response of the superconducting bridge to a supercritical current.

In the opposite limit the quasiparticles are thermalized and $f(\epsilon)$ can be described by the Fermi-Dirac distribution function
with a local temperature and chemical potential - so called quasi-equilibrium approach \cite{Nagaev,Giazotto}.
From Eqs. (2) and (4) one may obtain (see Appendix) the heat conductance equation for the local temperature of the quasiparticles
$T_{loc}=T+\delta T_{loc}$
\begin{equation} 
C_v\frac{\partial \delta T_{loc}}{\partial t}=\kappa \nabla^2
\delta T_{loc}+N_0\frac{T}{T_c}\frac{\partial|\Delta|^2}{\partial t}
-C_v\frac{\delta T_{loc}}{\tau_{in}}
\end{equation}
where $C_v=2\pi^2k_B^2N_0T/3$ is the electron heat capacity and $\kappa=2\pi^2k_B^2DN_0T/3$ is the electron heat conductivity in the normal state.
In this limit $\tau_{in}=\tau_{e-ph}$ or $\tau_{esc}$ whatever is larger. In Eq. (6) we neglect the heating
effects due to Joule dissipation which is valid for our problem near $T_c$ (for discussion see Appendix).

The time dependent equation for $\Delta$ in this limiting case resembles the ordinary time-dependent Ginzburg-Landau equation
with time and coordinate dependent local temperature
\begin{eqnarray}
\frac{\pi\hbar}{8k_BT_c} \left(\frac{\partial }{\partial
t}+2ie\varphi \right) \Delta=
\\
\nonumber =\xi_{GL}^2\left( \nabla -i\frac{2e}{\hbar c}A\right)^2\Delta
+\left(1-\frac{T_{loc}}{T_c}-\frac{|\Delta|^2}{\Delta_{GL}^2}\right)\Delta.
\end{eqnarray}
(in this form Eq. (7) was derived earlier in Ref. \cite{Galaiko}).

Eqs. (6-7) and $div j=0$ are the basic equations that govern the transient response of the superconducting bridge in the quasi-equilibrium approach.

In numerical calculations for the 1D case we assume that the superconducting bridge of finite length $L$ is attached to massive
superconducting electrodes which are being in equilibrium (physically it corresponds to the variable thickness bridge).
It imposes the following boundary conditions: $\Delta(x=0,L)=\Delta_{eq}exp(i\varphi(x=0,L)t)$, $T_{loc}(x=0,L)=T$, $\varphi(0)=0$,
$\varphi(L)=V$ and the voltage $V$ can be found from integrating Eq. (5) over the length of the bridge. For $f_L$ more complicated boundary
conditions are used: $f_L(x=0,L)=f_L^0$ when $\epsilon>\Delta_{eq}$ and $\partial f_L/\partial x(x=0,L)=0$ for smaller energies.

In the two-dimensional case we assume that the superconducting bridge of finite length $L$ and finite width $w$ is attached to massive normal electrodes
being in equilibrium. Normal electrodes considerably simplify our calculation in the current constant regime, which could be easily implemented
via boundary conditions for the electrostatic potential $-\sigma_n \partial \varphi/\partial x(x=0,L)=j$. The rest of the boundary conditions are
as follows: in longitudinal (x) direction - $\Delta(x=0,L)=0$, $T_{loc}(x=0,L)=T$, $f_L(x=0,L)=f_L^0$, and in transverse (y) direction we use
ordinary superconductor-isolator boundary conditions. To diminish the influence of nonequilibrium effects from NS boundaries we locally increase
$T_c$ near the ends of the bridge (on the distance $5\xi_0$ from each end) by 20 $\%$.

In our calculations we use the following natural variables as the units of the corresponding quantities  :
$t_0=\hbar/\Delta_0$, $\Delta_0=1.76k_BT_c$,  $\xi_0=\sqrt{\hbar D/\Delta_0}$, $q_s^0=\hbar c/2e\xi_0$,
$j_0=\Delta_0\sigma_n/(\xi_0e)$ and $\varphi_0=\Delta_0/e$. In our calculations we
use $\tau_{in}/t_0=8-1000$ which are typical values for many superconductors (for example in $MgB_2$ $\tau_{e-ph}/t_0 \simeq 20$, in Nb
$\tau_{e-ph}/t_0 \simeq 100$, in Sn $\tau_{e-ph}/t_0 \simeq 200$ and in $Al$ $\tau_{e-ph}/t_0 \simeq 1000$).

In our numerical simulations we assume that the current increases linearly (from $t=0$) during the time interval $\delta t=5t_0$. Such a procedure
provides a better numerical stability of our calculations in comparison with an instant application of the current. Time delay depends weakly
on the specific choice of $\delta t$ while $t_d\gg \delta t$.

\section{Transient response in 1D case}

\subsection{Non-thermal model}

For simplicity we first neglect the nonequilibrium contribution $\delta j_s$ to the supercurrent. Effect of finite $\delta j_s$ on $t_d$ will
be discussed at the end of this subsection.

In the spatially uniform case one may find from Eqs. (2,3) the equation for the dynamics of the dimensionless magnitude of the order parameter
$f=|\Delta|/\Delta_{eq}$

\begin{equation}
\tau_{GL}\frac{\partial f}{\partial t}+ afY(f,T)\int_{0}^{t}\frac{\partial f}{\partial t'}e^{-(t-t')/\tau_{in}}dt'
=f\left(1-f^2-\frac{\tilde{I_s}^2}{f^4}\right),
\end{equation}
where $a=\pi\Delta_{GL}/4k_BT(1-T/T_c)^{1/2}\simeq 2.4/(1-T/T_c)^{1/2}$, $\tau_{GL}=\pi \hbar/8k_B(T-T_c)$,
\begin{equation}
Y(f,T)=\frac{2}{\pi}\int_{1}^{\infty} \frac{d\epsilon} {\epsilon (\epsilon^2-1)^{1/2} {\rm cosh}(\epsilon f \Delta_{eq}/2k_BT)},
\end{equation}
and $\tilde{I_s}^2=4/27 (I_s/I_{dep})^2$ ($I_{dep}=j_{dep}wd$ is the Ginzburg-Landau depairing current, $I_s=j_swd$ and we use
the initial condition $f_L=f_L^0$ at $t=0$ when the current is turned on). In case when $\Delta_{eq}/2k_BT \leq 1/2$ one has
$Y(f,T)\simeq 1-bf$ ($b=\Delta_{eq}/2k_BT\simeq 1.52(1-T/T_c)^{1/2}$). Note that in Refs. \cite{Tinkham_review,Geier} $Y(f)=1$,
which is valid when $b\ll 1$. But $b\simeq 0.48$ already at $T=0.9T_c$ and the term $-bf$ in $Y(f)$ should be taken into account.

In Eq. (8) one may neglect the first term on the left hand side (LHS) in comparison with the second term when $\tau_{GL}\ll \tau_{in}/(1-T/T_c)^{1/2}$
(except the very beginning of the transition period when the time integral in Eq. (8) is small). With this simplification from Eq. (8) one may
find the differential equation for $f$
\begin{widetext}
\begin{equation}
\tau_{in}\frac{\partial f}{\partial t}=\frac{1-f^2-\tilde{I_s}^2/f^4}{aY(f,T)+2f-4\tilde{I_s}^2/f^4+Y'(f,T)(1-f^2-\tilde{I_s}^2/f^4)/Y(f,T))}.
\end{equation}
\end{widetext}

In the limit $Y(f,T)=1$ Eq. (10) coincides with Eq. (64) of Ref. \cite{Tinkham_review}. Eq. (10) becomes invalid at $f=f_{min}$ when the denominator goes
to zero
\begin{widetext}
\begin{equation}
aY(f_{min})+2f_{min}-4\tilde{I_s}^2/f_{min}^4+Y'(f_{min})(1-f_{min}^2-\tilde{I_s}^2/f_{min}^4)/Y(f_{min})=0,
\end{equation}
\end{widetext}
and in the beginning of the transition period when $f$ changes on a time scale $\sim \tau_{GL}$ from 1 to $f_{max}<1$.
$f_{max}$ could be found from the following equation
\begin{equation}
\tilde{I_s}^2=(1-f_{max})f_{max}^4(a(Y(f_{max})+1+f_{max})
\end{equation}
which results from Eq. (8) if one assumes a step like (on a time scale $\tau_{GL}\ll \tau_{in}$) decrease of $f$ from 1 to $f_{max}$
(Eq. (12) transits to Eq. (62) of Ref. \cite{Tinkham_review} when $Y(f_{max})=1$).

From Eqs. (10-12) one can find the delay time
\begin{widetext}
\begin{equation}
\frac{t_d}{\tau_{in}}=\int_{f_{max}}^{f_{min}}\frac{aY(f,T)+2f-4\tilde{I}^2/f^4+Y'(f,T)(1-f^2-\tilde{I}^2/f^4)/Y(f,T)}{1-f^2-\tilde{I}^2/f^4}df
\end{equation}
\end{widetext}
where we replace the superconducting current $I_s$ by the full current $I$ because when $f$ changes from $f_{max}$ to $f_{min}$ the normal current
in the wire is much smaller than $I$ and $I_s\simeq I$.

\begin{figure}[hbtp]
\includegraphics[width=0.53\textwidth]{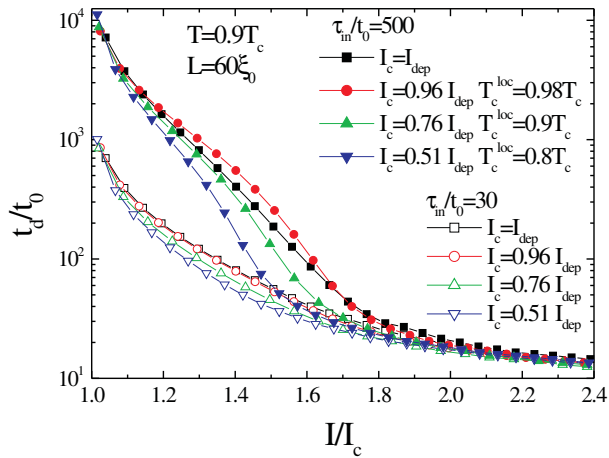}
\caption{Dependence of the time delay $t_d$ on the normalized current $I/I_c$ in the superconducting bridge with locally suppressed $T^{loc}_c$
(in the area with size $\xi_0$ in the center of the bridge) and two values of $\tau_{in}$. Decrease of $t_d$ with decreasing $T^{loc}_c$ is explained
by the locally smaller value of $|\Delta|$ in equilibrium and it takes less time to suppress it to zero. In the case of a defect-free bridge ($I_c=I_{dep}$)
the time delay is smaller than in the bridge with a weak defect (at the same value of normalized current $I/I_c$) because $|\Delta|\to 0$ near the ends
of the homogenous bridge where the cooling effect is weaker due to the diffusion of nonequilibrium quasiparticles to the leads.}
\end{figure}

In the spatially nonuniform case we solve the set of Eqs. (2-5) numerically (in Eq. (5) we put $\delta j_s=0$)
and we find that even for a homogenous bridge, $|\Delta|$ decays faster near the ends of the bridge. At first sight this result looks rather
unexpected, because at the ends $|\Delta|$ is maximal due to the boundary conditions (which originate from the proximity with the
massive superconducting leads being in equilibrium). But effective cooling of quasiparticles is weaker near the ends of the bridge, because
nonequilibrium quasiparticles with energy $\epsilon>\Delta_{eq}$ can freely diffuse
away from the bridge to the leads. As a result $|\Delta|$ decreases faster near the ends of the bridge.

We checked that a similar effect exists (see also Ref. \cite{Oppenheim}) even in the so called local equilibrium limit (when
$L_{in}=(D\tau_{in})^{1/2}\ll \xi(T)$ \cite{Kramer1,Watts-Tobin}) and one can neglect the diffusion of quasiparticles.
But in that case the spatial gradient of $|\Delta|$ along the bridge is considerably smaller and the effect appears
only in some range of currents and $\tau_{in}$. This result shows that not only diffusion of nonequilibrium quasiparticles may provide this
effect but also the gradient of $|\Delta|$ which appears near the ends of the variable thickness bridge when $I \neq 0$.
\begin{figure}[hbtp]
\includegraphics[width=0.53\textwidth]{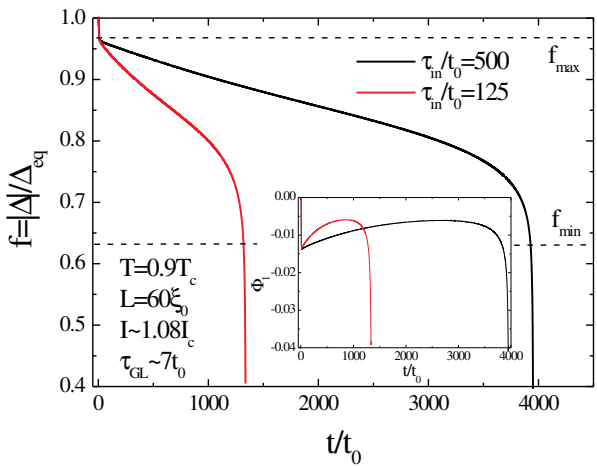}
\caption{Time evolution of normalized $|\Delta|$ in the center of the bridge with a weak defect after the application of the supercritical
current at $t=0$. Dashed lines shows values of $f_{max}$ and $f_{min}$ from Eqs. (11,12).
In the inset we show the time dependence of $\Phi_1$ in the center of the bridge. Dynamics of both $|\Delta|$ and $\Phi_1$ does not depend on
$\tau_{in}$ in the beginning and in the end of the transition process.}
\end{figure}

Real bridges are never homogenous. Variations of their physical (mean free path length and/or $T_c$) and geometrical
(width or thickness) properties along the superconductor may exist. For example we find that even a $2\%$ suppression of $T_c$
on a length scale of $\xi_0$ in the center of the bridge favors the local suppression of $|\Delta|$ in comparison with its suppression near the ends of the bridge
(in the studied temperature interval $0.9-0.98 T_c$). We checked that the time delay $t_d$ varies with a change of local $T_c$ (see Fig. 1)
but the functional dependence $t_d(I/I_c)$ stays practically the same when the suppression of $I_c$ due to defects is not strong ($I_c\simeq I_{dep}$).
Further we consider the superconducting bridge with a defect in the center where $T_c$ is locally suppressed by 2$\%$.

In Fig. 2 we present the time evolution of the order parameter in the center of the superconducting bridge with
length $60 \xi_0$ at $T=0.9 T_c$ and two values of $\tau_{in}$. One can see that qualitatively the dynamics of $|\Delta|$ follows the
predictions of the spatially uniform model. At the beginning of the transition there is a sudden drop in $|\Delta|$ (on a time scale $\sim \tau_{GL}$) and
the value of this drop is close to the one predicted by Eq. (12) (note that it does not depend on $\tau_{in}$). Further
decay of $|\Delta|$ strongly depends on $\tau_{in}$ (see Fig. 2) until it reaches some minimal value which is close to the one predicted by
Eq. (11). After that $|\Delta|$ varies fast in time (again on the time scale $\sim \tau_{GL}$) and its dynamics weakly depends on $\tau_{in}$.

In Fig. 3 we present the dependence $t_d(I/I_c)$ for different values of $\tau_{in}$. If one compares this result with Eq. (13)
(solid curve in the inset of Fig. 3) one can see that for the spatially nonuniform case the time delay is much shorter
(when $\tau_{in} \gg \tau_{GL}$). Besides there is
no 'universal' curve (see the inset to Fig. 3 for normalized variables) on which all curves drop as in the case of the spatially uniform model.
We explain these deviations by the diffusion of nonequilibrium quasiparticles from the point where $|\Delta|$ decreases faster. Indeed, the
diffusion decreases locally the cooling of quasiparticles and accelerates the decay of $|\Delta|$. This effect also results in a dependence
of $t_d$ on the position, where $|\Delta|$ locally decays, with respect to the ends of the bridge (compare $t_d$ for an uniform bridge
and for the bridge with a weak defect in Fig. 1) and on the length of the bridge when $L/2 \lesssim L_{in}=(D\tau_{in})^{1/2}$ (see Fig. 4).
\begin{figure}[hbtp]
\includegraphics[width=0.48\textwidth]{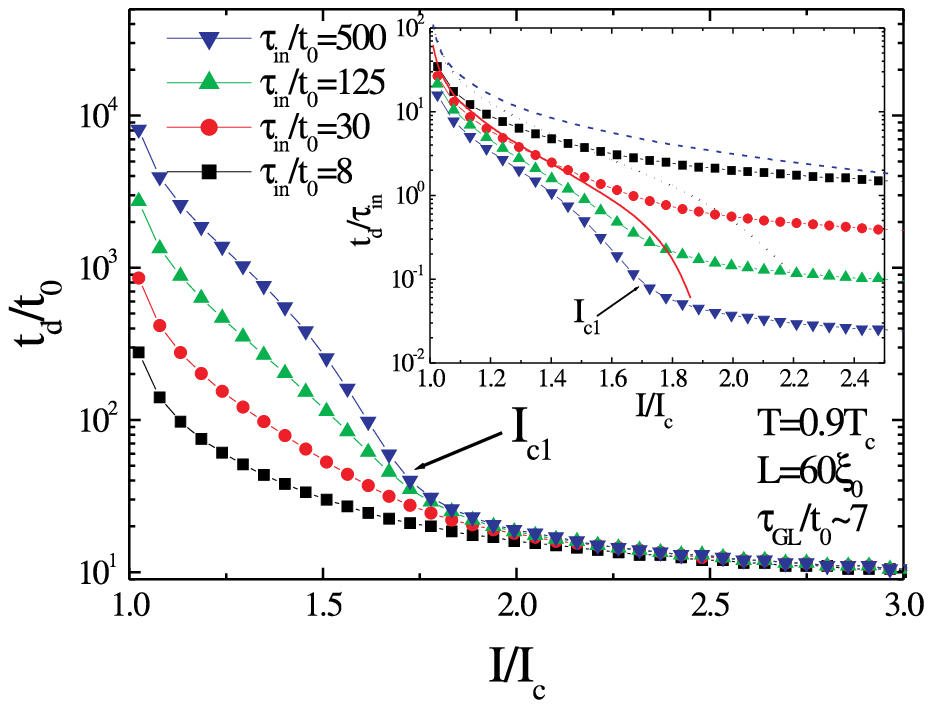}
\caption{Dependence of the time delay on the normalized current for different $\tau_{in}$ (both comparable and much larger than $\tau_{GL}$).
In the inset we show the same dependencies but with $t_d$ normalized in units of $\tau_{in}$. Solid curve corresponds to Eq. (13), dotted
curve to Eq. (13) with $Y(f)=1$ (Eq. (64) in Ref. \cite{Tinkham_review}) and dashed curve from Eq. (60) of Ref. \cite{Tinkham_review}
(local equilibrium limit).}
\end{figure}

In the model described by Eq. (13) $t_d \to 0$ when $f_{min}=f_{max}$ which occurs for some current $I_{c1}$ as given in Ref. \cite{Tinkham_review}.
Spatially nonuniform decay of $|\Delta|$ does not change this result qualitatively - the time delay drops fast when $I\to I_{c1}$
(but its value is larger in the uniform model - see inset in Fig. 3) while at larger currents $t_d$ is still finite and it does not depend
on $\tau_{in}$ anymore. From a physical point of view for currents $I>I_{c1}$ 'cooling' of quasiparticles (which is limited by a decrease of $|\Delta|$
from $\Delta_{eq}$ to zero) cannot compensate the depairing effect of the current and the time delay does not depend on $\tau_{in}$.
Finite delay time at $I>I_{c1}$ is explained by the finite time $\tau_J=\tau_{GL}/u$ ($u\sim \pi^4/14\zeta(3)$) \cite{Ivlev}
needed for the transformation of the normal current to the superconducting one (assuming that the normal current appears in the bridge
on a time scale much smaller than $\tau_J$) and finite relaxation time of $|\Delta|$, which imposes a relatively weak dependence on the
current $t_d(I)\sim \tau_{GL}I_c/I$.
\begin{figure}[hbtp]
\includegraphics[width=0.48\textwidth]{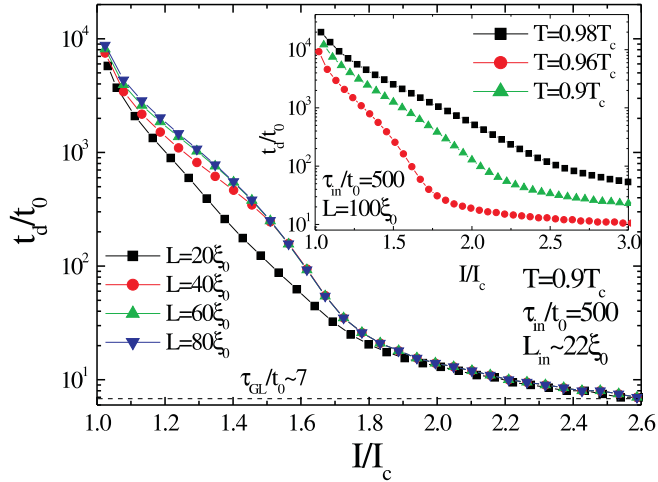}
\caption{Dependence of the time delay on the normalized current for bridges with different length. For chosen $\tau_{in}/t_0=500$ and
$L_{in}/\xi_0\simeq 22\xi_0$ there is a strong decrease of $t_d$ when $L\lesssim 2L_{in}\sim 40 \xi_0$
which we explain by increased diffusion of nonequilibrium quasiparticles to the leads. In the inset we show the dependence of $t_d$ on temperature
for a long bridge $L\simeq 4.5L_{in}$.}
\end{figure}

Notice that not only diffusion of nonequilibrium quasiparticles is responsible for the reduction of $t_d$. It turns out that if we remove
the term proportional to $q_s^2$ in Eq. (4) then the time delay and the current $I_{c1}$ increases (but $t_d$ will still be smaller than in the
uniform case). Finite $q_s$ smears out the peak in the density of states $N_1$ and in the spectral function $R_2$ at $\epsilon=|\Delta|$
\cite{Anthore} which results in a somewhat smaller value of $\Phi_1$, in comparison with the case when $q_s=0$, and thus smaller cooling
effect.

Let us now discuss how the nonequilibrium contribution to the supercurrent $\delta j_s$ influences the time delay. From Fig. 5 one can see that
finite $\delta j_s \neq 0$ leads to larger $t_d$ and $I_{c1}$. In some respect finite $\delta j_s$ 'compensates'
the reduction of $t_d$ due to diffusion of quasiparticles and smearing of the spectral functions near $\epsilon=|\Delta|$ and shifts
$t_d$ closer to the result expected from the uniform model. Originally the increase of $t_d$ due to finite $\delta j_s$ was predicted in Ref. \cite{Geier}
but authors used spatially uniform model and their values of $t_d$ was larger than Tinkham's result with $\delta j_s=0$
(see Fig. 5 in Ref. \cite{Tinkham_review}) which is shown in Fig. 5 by the dashed curve.
\begin{figure}[hbtp]
\includegraphics[width=0.48\textwidth]{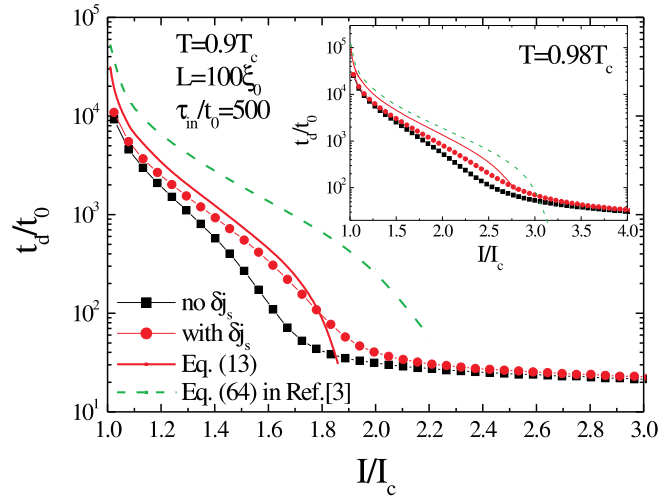}
\caption{Change in the time delay when one takes into account the nonequilibrium contribution to the superconducting current.
From the inset one can see that at higher temperature the effect of $\delta j_s \neq 0$ becomes smaller.}
\end{figure}

\subsection{Quasi-equilibrium model}

In the uniform case, from Eqs. (6,7) follows the equation for the time dependence of $f=|\Delta|/\Delta_{eq}$ which is similar to Eq. (8)
\begin{equation}
\tau_{GL}\frac{\partial f}{\partial t}+ a_Tf\int_{0}^{t}\frac{\partial f^2}{\partial t'}e^{-(t-t')/\tau_{in}}dt'
=f\left(1-f^2-\frac{\tilde{I_s}^2}{f^4} \right)
\end{equation}
where $a_T=3\Delta_{GL}^2/2\pi^2k_B^2T_c^2\simeq1.42$. Again, when $\tau_{GL} \ll \tau_{in}$ we neglect the first term on the LHS of Eq. (14)
and we find the differential equation
\begin{equation}
\tau_{in}\frac{\partial f}{\partial t}=\frac{1-f^2-\tilde{I_s}^2/f^4}{2f(1+a_T)-4\tilde{I_s}^2/f^5}.
\end{equation}

Similar to the case which is considered in section IIIA, Eq. (15) is not valid at the very beginning and at the end of the transition period when
the denominator of Eq. (15) goes to zero. From Eq. (15) one may find the time delay
\begin{equation}
\frac{t_d}{\tau_{in}}=\int_{f_{max}}^{f_{min}}\frac{2f(1+a_T)-4\tilde{I}^2/f^5}{1-f^2-\tilde{I}^2/f^4}df,
\end{equation}
where $f_{min}=(2\tilde{I}^2/(1+a_T))^{1/6}$ and $f_{max}$ should be found from the equation
\begin{equation}
\tilde{I}^2=f_{max}^4(1+a_T)(1-f_{max}^2).
\end{equation}

From Eq. (16) it follows that $t_d$ depends only on temperature via $\tau_{in}(T)$. This is the main qualitative difference with Eq. (13),
where the strong temperature dependence of $t_d$ comes from $a(T)$ and $Y(T)$. As a consequence, in the quasi-equilibrium model the current $I_{c1}
\simeq 1.55 I_c$ (when $f_{max}=f_{min}$ and $t_d\to 0$).

In Fig. 6 we present the dependence $t_d(I/I_c)$ for a bridge with length $L=100 \xi_0$ found from a numerical solution of Eqs. (6-7) and $div j=0$.
As in the non-thermal model the effect of quasiparticle diffusion is large and in the homogenous bridge the order parameter decays faster near the leads.
Therefore we suppress $T_c$ in the center of the bridge by $4 \%$ in the region with size $\xi_0$ and in Fig. 6 we show the time decay of
the order parameter in this place. In the inset one can see rather weak temperature dependence of $t_d$ at currents near $I_c$ and
strong temperature dependence at $I \gtrsim I_{c1}\sim 1.55 I_c$. The later occurs due to the strong temperature dependence of $\tau_{GL}$ which
mainly determines the time delay at $I\gtrsim I_{c1}$.
\begin{figure}[hbtp]
\includegraphics[width=0.53\textwidth]{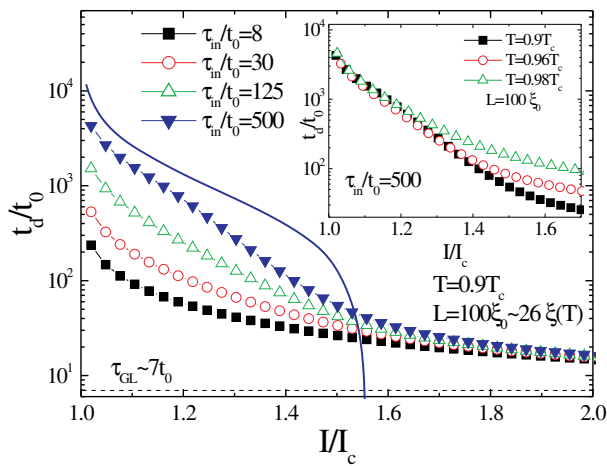}
\caption{Dependence of the time delay on the normalized current for a superconducting bridge (quasi-equilibrium model).
Solid curve corresponds to Eq. (16) with $\tau_{in}/t_0=500$. In the inset we show the dependence $t_d(I/I_c)$ for different temperatures.}
\end{figure}

From a comparison of Figs. 3 and 6 one can see that the dependence of the time delay on current and $\tau_{in}$ in the quasi-equilibrium limit
resembles the one in the non-thermal limit but it is shorter for fixed $I/I_c$ and $\tau_{in}$. We can explain this as follows.
In the non-thermal model the cooling effect comes from the term $R_2\partial |\Delta|/\partial t$ in the kinetic equation which is peaked at
energies $\epsilon \simeq |\Delta|$ and it has a long tail $\sim |\Delta|/\epsilon$ at larger energies. It results in the largest deviation from
equilibrium at $\epsilon \simeq |\Delta|$ and this interval of energies gives a relatively large contribution to $\Phi_1$. In the quasi-equilibrium model,
due to the thermalization process this peak is smeared out and it leads to a smaller value of $\Phi_1$ and a weaker 'cooling' effect.

\section{Dynamic response in 2D bridge}

Now we study the dynamic response in a two dimensional bridge with nonuniform current distribution across the superconductor. In our case the
nonuniformity arises due to the application of a perpendicular magnetic field (see Fig. 7). We consider only weak fields when there is no vortices
in the bridge at currents below $I_c$.
\begin{figure}[hbtp]
\includegraphics[width=0.53\textwidth]{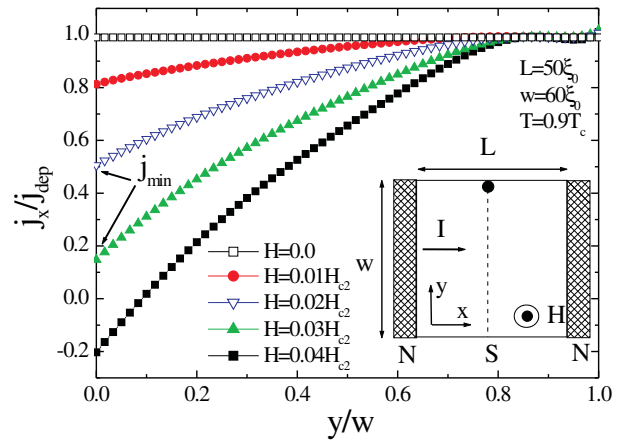}
\caption{Distribution of the current density across the superconducting bridge
(along the dashed line in the sketch of the bridge shown in the inset) at different magnetic fields and currents just below $I_c(H)$.
In the inset we show a sketch of the bridge contacted to normal leads. To prevent the influence of nonequilibrium effects from
the NS boundaries we locally (on the distance 5$\xi_0$ from each end) increased the local $T_c$ by 20 $\%$.}
\end{figure}

Due to the nonuniform current distribution the order parameter decays first near the edge where the current density is maximal. When near the edge
$|\Delta|\to 0$ the vortex enters the film and passes through it (for our length of the bridge only one vortex enters simultaneously).
Behind the moving vortex there is a wake - region with suppressed order parameter (see inset (b) in Fig. 8) which appears due to the large recovery
time of $|\Delta|$ (in the local equilibrium approximation, when $\tau_{in}<\tau_{GL}$, this effect was studied
in Refs. \cite{Glazman,Vodolazov_lattice}). This wake favors faster nucleation of the second and subsequent vortices
(in Fig. 8 every minimum in $|\Delta|(t)$ corresponds to the entrance of a new vortex) because of subsequent gradual suppression of $|\Delta|$
(see Fig. 8). Passage of several vortices nucleates a quasi-phase slip line \cite{Vodolazov_lattice}  - the region with width $\sim 2\xi$
where $|\Delta|$ is strongly suppressed but is still different from zero
(see insets (c,d) in Fig. 8). This quasi-phase slip line (PSL) may convert or not to a normal domain which than spreads over the superconductor
if the current or $\tau_{in}$ is large enough. Note that in the present model only partial Joule dissipation is considered
(via the dependence of the Green functions on the supervelocity - see Ref. \cite{Vodolazov_heating}) and hence the time for nucleation of the
quasi-PSL is underestimated.

When the ratio $j_{min}/j_{dep}$ is small (for definition of $j_{min}$ see Fig. 7) the quasi PSL may not appear at $I\sim I_c(H)$
because in the part of the film where $j$ is small the vortices move slowly and $|\Delta|$ has time to recover. Does quasi-PSL appear
or not at $I=I_c(H)$
is controlled by $\tau_{in}$ - the larger $\tau_{in}$ the smaller the current threshold value $I_{th}(\tau_{in})$ when quasi-PSL appears in the bridge
\cite{Vodolazov_lattice}. If $I_{th}>I_c(H)$ then in the current interval $I_c(H)<I<I_{th}$ there is slow vortex motion and the
voltage signal is small while for $I>I_{th}$ there will be a sudden jump in the voltage connected with the appearance of the quasi-PSL.
\begin{figure}[hbtp]
\includegraphics[width=0.48\textwidth]{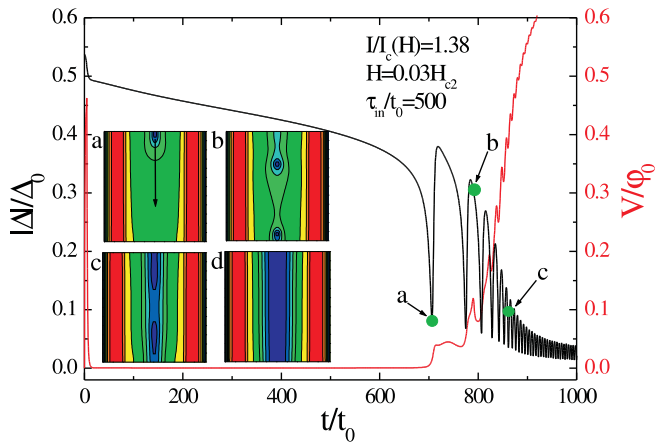}
\caption{Time dependence of $|\Delta|$ at the edge of the bridge (marked as black spot in the inset in Fig. 7) and voltage drop across the bridge at
$I/I_c(H)\simeq 1.38$ and $H=0.03 H_{c2}$ (non-thermal model). In the inset we present snap-shots of $|\Delta|$ in the bridge at different
moments in time: a) $t/t_0=706$, b) $t/t_0=789$, c) $t/t_0=863$ and d)- $t/t_0=1500$. Arrow in inset a) shows the direction of vortex motion.
The narrow peak in the voltage at $t\simeq 0$ is connected with initially normal current $I_n=I$ which transforms to the superconducting one
on the time scale $\tau_J$.}
\end{figure}

Below we consider the situation when $I_c(H)>I_{th}$. Due to the nonuniform (over width of the bridge) decay of $|\Delta|$ there is an uncertainty in
the definition of the time delay. One of the variants is to define it as the time needed for the nucleation of the first vortex after the
application of the supercritical current (this time is shown in Fig. 9). From Fig. 8 it is clear that this definition underestimates $t_d$
because a large voltage response appears only after the nucleation of the quasi-PSL. But noticeable difference arises only at relatively large
currents when these two times becomes comparable with each other (for example for the parameters of Fig. 8 the first vortex nucleation time
is $\sim 700 t_0$ while the PSL nucleation time is $\sim 900 t_0$).
\begin{figure}[hbtp]
\includegraphics[width=0.48\textwidth]{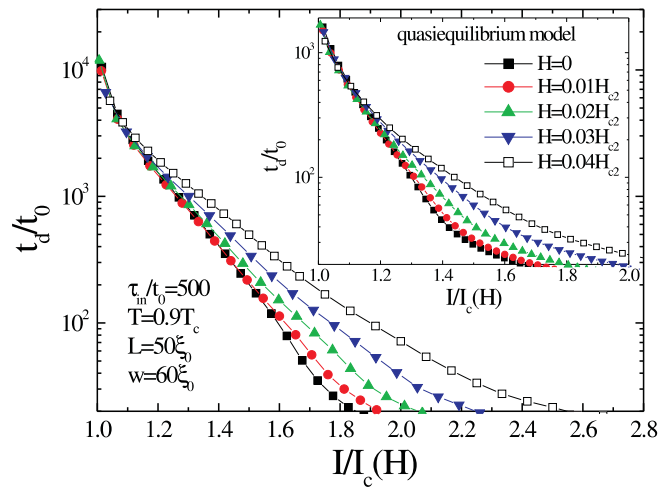}
\caption{Dependence of the time delay (in the appearance of the first vortex) on the applied current at different
magnetic fields (non-thermal model). In the inset we show the same dependencies calculated in the quasi-equilibrium model.}
\end{figure}

Our simulations show that the delay in the appearance of the quasi-phase slip line after the nucleation of the first vortex is mainly determined
by the 'flight' time of the first vortex across the bridge. We find that the second, third and so on
vortices enter the bridge before the first vortex exits. The number of vortices simultaneously present in the bridge
depends on the width of the bridge, $\tau_{in}$ and on the ratio $j_{min}/j_{dep}$. For example in the bridge with $w=50 \xi_0$
there are two vortices in the quasi-phase slip line (see insets b,c in Fig. 8), while in the bridge with $w=120 \xi_0$ there are
already four vortices (for the same $\tau_{in}$ and ratio $j_{min}/j_{dep}$).

The 'flight' time of the first vortex $t_f$ could be roughly estimated using results for the time delay in quasi-1D bridges and the knowledge about
the basic properties of the vortex. First of all one should remind that next to the vortex core the superconducitng current density $\sim j_{dep}$.
From one side of the vortex this current density and the transport current density are summed up and it leads to a local destruction
of superconductivity and vortex motion in that direction. If the sum of these current densities exceeds $j_{c1}=I_{c1}/wd$ then the time decay of $|\Delta|$ is rather
short ($t_d\sim \tau_{GL} I_{dep}/I$ - see section III) and the first vortex moves fast. One can estimate its average velocity as
$v_{aver}\sim \xi(T)/t_d$. For our bridge with $w=60 \xi_0\simeq 16 \xi(T=0.9T_c)$ the 'flight' time of the first vortex is about
$t_f\simeq w/v_{aver}\sim 16 \tau_{GL} \sim 112 t_0$ (for parameters of Fig. 8 where $I_c(H=0.03H_{c2})=0.75 I_{dep}$)
which is not far from the numerical value $\sim 83 t_0$.

As the current approaches $I_c$ the nucleation time of the first vortex increases much faster than the first vortex 'flight' time
and the time delay is mainly determined by the former time (except for very wide films with
$w \gg \xi t_d/\tau_{GL}\sim \xi \tau_{in}/\tau_{GL}$).

The above rough estimations are valid if the sum of the current densities from the vortex and the transport current exceeds $j_{c1}$.
In case of a small ratio $j_{min}/j_{dep}$ this condition is not fulfilled and the first vortex moves with a much lower velocity. As a limiting
case the quasi PSL is not nucleated and at $I\sim I_c$ there is only slow vortex motion.

In the quasi-equilibrium limit we have qualitatively the same results for the dependence of $t_d$ (time nucleation of the first vortex) on the applied magnetic
field - see inset in Fig. 9. Note that in both models at $I\simeq I_c(H)$ this time delay slightly decreases with an increase of $H$
(probably it is connected with a stronger diffusion of the nonequilibrium quasiparticles in the 2D case in comparison with 1D) while at larger currents $t_d$
increases (for the same ratio $I/I_c(H)$). The last effect can be connected with the current redistribution when $|\Delta|$ becomes suppressed
near one edge and the superconducting current escapes that region which leads to a locally smaller value of the current density.

\section{Dynamic response on an alternating current pulse}

In this section we study the dynamic response of the superconducting bridge on an alternating current pulse with amplitude $I_{ac}$ larger than $I_c$
and with a zero time average (the ac current pulse is modelled as one period of the sinusoid $I=I_{ac}sin(2\pi t/T_{ac})$). Our interest to this problem
arises from recent work \cite{Probst} where the voltage response of the superconducting YBCO bridge on a short pulse of synchrotron radiation
(with duration of several picoseconds) was experimentally observed  even in the absence of dc current. Due to the absence of a bolometric
origin of the resistive response authors supposed that the electric field of the electromagnetic
radiation accelerates the superconducting electrons and when the radiation induced current exceeds the critical current a finite voltage appears
in the bridge.
\begin{figure}[hbtp]
\includegraphics[width=0.48\textwidth]{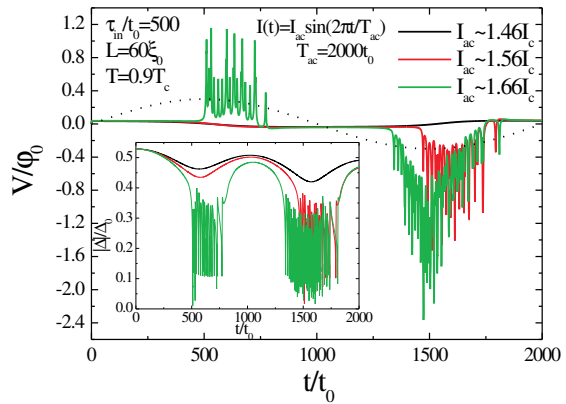}
\caption{Time dependence of the voltage in the 2D superconducting bridge during an ac current pulse (marked as dotted line).
In a relatively narrow range of amplitudes of the current pulse the larger voltage appears only for one direction of the current.
In the inset we show the time dependence of the order parameter in the center of the bridge. Calculations are made within the non-thermal model.}
\end{figure}

Based on the results of sections III-IV we may give the following rough criterion: at given amplitude $I_{ac}$ a resistive response does exist
when during the ac current pulse the time interval when $|I|>I_c$ exceeds $t_d(I_{ac})$, where $t_d(I_{ac})$ corresponds to the time delay
on the abrupt switching on of the dc current with amplitude $I_{ac}$.
We numerically checked and confirmed this idea on the example of a quasi-1D superconducting bridge. We also find one interesting effect which arises when
the amplitude of the ac pulse approaches some critical value. From Fig. 10 one can see that
with increasing $I_{ac}$ a large voltage appears first in the second half of the ac current
pulse and one needs to increase $I_{ac}$ to observe it in the first half too. We explain this effect as follows. During the first half of
the pulse the order parameter is getting suppressed and it does not recover its equilibrium value at $t=T/2$ (when $I(t)=0$) due to
the finite relaxation time of $|\Delta|$ (see inset in Fig. 10 for $I=1.46 I_c$ and $I=1.56I_c$). At $T/2<t<T$ there is further suppression
of $|\Delta|$ and it goes to zero (when $I_{ac}$ is sufficiently strong) and a highly resistive state appears in the bridge (in the used 1D bridge
it is realized as a phase slip process). As a result the time-averaged voltage is not equal to zero and its sign depends on the phase of the ac
current pulse - with change of the phase by $\pi$ the sign of the time averaged voltage changes.

Dynamic response of a 2D bridge with nonuniform current distribution is qualitatively similar to the case of a 1D bridge. When the amplitude of
the ac pulse exceeds some critical value (at fixed period $T_{ac}$) the vortices enters the bridge one by one and they suppress $|\Delta|$ along
their trajectory of motion (see Fig. 11).
At $t=T/2$ there are no vortices in the bridge (for the chosen parameters in Fig. 11) but $|\Delta|$ is suppressed below its
equilibrium value and it facilitates the
faster vortex motion in the second half of the pulse. As a result the voltage is larger in the second half of the pulse (see Fig. 11).
\begin{figure}[hbtp]
\includegraphics[width=0.48\textwidth]{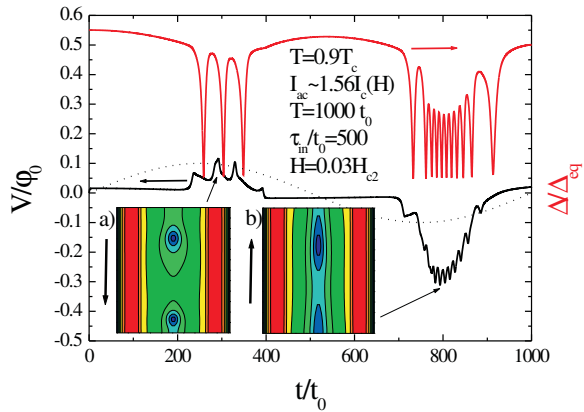}
\caption{Time dependence of the voltage across the bridge and magnitude of order parameter in the center of superconductor when ac current pulse
(marked as dotted line) is applied. In the inset we show spatial distribution of $|\Delta|$ at $t=290 t_0$ (a) and at $t=800 t_0$. In the second half
of the pulse the order parameter is suppressed strongly along the path of vortex motion which ensures larger vortex velocity and
higher value of the voltage than in the first half of the current pulse. Calculations are made in quasi-equilibrium model.}
\end{figure}

\section{Discussion}

In the first experiment on transient response it was found that the time delay in Al films does not depend on temperature when
$0.76<T/T_c<0.92$ \cite{Pals}. Subsequent experiments on Al \cite{Wolter,Attekum,Wolter2} and In \cite{Frank} films found a strong
temperature dependence of $t_d$ near $T_c$. A later experiment on a YBCO bridge again did not reveal any temperature dependence of $t_d$
in a wide temperature interval $4.2K-68K$ \cite{Jelila}. Note that all previous theories predicted a strong temperature dependence of
$t_d$ near $T_c$ \cite{Tinkham_review,Geier} due to the coefficient $a(T)$ in Eq. (13). In Ref. \cite{Jelila} authors tried to resolve
this problem by the replacement the product $a(T)\tau_{in}\sim \tau_{|\Delta|}$ by a temperature independent escape time of nonequilibrium
phonons $\tau_{esc}$ to the substrate. Our results give a physical and mathematical reasons for such a replacement.
In case if the escape time of phonons to the substrate is the longest relaxation time the electrons and phonons have one temperature
which is different from the bath temperature.
In this case one should use the quasi-equilibrium model with $\tau_{in}=\tau_{esc}$ and this model predicts a temperature independent
$t_d(I/I_c(T))$ at currents relatively close to $I_c$ (see inset in Fig. 6 and Eq. (16)).

In the case when there is good thermal connection between the superconductor and the substrate and $\tau_{esc}\ll \tau_{e-ph}\ll \tau_{e-e}$
the non-thermal model is more relevant. This model predicts strong temperature dependence of $t_d$ near $T_c$ (see inset in Fig. 4) which
is even stronger than that follows from Refs. \cite{Tinkham_review,Geier} because of the temperature dependent $Y(T)$.
In a more complicated situation the times $\tau_{e-ph}$, $\tau_{e-e}$ and $\tau_{esc}$ could be comparable with each other which brings a
more complex temperature dependence of $t_d(T)$ (it should be something in between two limiting cases considered here).

In our calculations we neglect the temperature dependence of $\tau_{in}$. It is approximately valid when $\tau_{in}=\tau_{esc}$ and not in the case when
$\tau_{in}=\tau_{e-ph}\sim 1/T^3$ (this temperature dependence comes from the Debye model of phonons). However if one does not go far from $T_c$
it gives only small corrections in $\tau_{e-ph}$ (for example $\tau_{e-ph}(0.9T_c)\simeq 1.4 \tau_{e-ph}(T_c)$).

How is the time delay  modified when we consider low temperatures? At low temperatures
there is an exponential decay of $Y(T)$ when $k_BT\ll \Delta_{eq}$ (which follows from the general expression for $Y(T)$ - see Eq. (9))
and it formally leads to a fast decay of $t_d$ when $T\to 0$ despite the increase of $\tau_{in}=\tau_{e-ph}$ (in non-thermal model
where the phonons are in equilibrium). But one should remember that Eqs. (2-5) were derived at $T\sim T_c$ (when $\Delta_{eq}\ll k_BT$)
and many terms were omitted, which can become important at low temperatures. Therefore this question needs additional investigation both
in non-thermal and quasi-equilibrium limits.

Our predictions for the dependence of the time delay of narrow superconducitng bridges/films (with width $\xi \ll w\ll \lambda^2/d$) on
the applied magnetic field could be checked by an experiment. We predict that in relatively weak magnetic fields $0<H \lesssim \Phi_0/4\pi \xi w$
(when the superconductor is in the Meissner state) the time delay depends weakly on $H$ for $I\gtrsim I_c(H)$
and for larger currents $I\gg I_c(H)$, $t_d$ should increase with increasing $H$. The crucial effect for the observation of long delay
times at $I\sim I_c(H)$ is the absence of trapped vortices in the superconductor. Indeed, our calculations show that for not very wide films $w \ll \xi
t_d/\tau_{GL}\sim \xi \tau_{in}/\tau_{GL}$ the time delay is determined mainly by the nucleation time of the first vortex and
the presence of trapped vortices should decrease $t_d$.

Our theoretical results for the dynamic response of a superconducting bridge on the short ac current pulse qualitatively resembles some of the
experimental results of Ref. \cite{Probst}. Namely we also find nonzero voltage response in the absence of a dc current and a change of its sign
when the phase of the ac current changes by $\pi$. However we use a very simple shape of the ac current pulse (sinusoid) which is drastically different
from the asymmetric pulse in the experiment (see Fig. 9 in Ref. \cite{Probst}). If the pulse is asymmetric (amplitude of
the current of one sign is much larger than the amplitude of the current of opposite sign) then the sign of the voltage response is determined
by the largest current in the pulse. This makes it difficult to directly compare our results with those of Ref.\cite{Probst} but in any case
we predict that the response may appear only if the duration of the pulse is larger than the time delay at the largest amplitude of the current pulse.

In Ref. \cite{Probst} no external magnetic field was present but the current distribution probably was nonuniform across the bridge
both due to current-crowding effect at the ends of the bridge \cite{Clem_crowding} and the small aspect ratio $L/w\sim 0.44$ \cite{Aslamazov}.
Indirectly it could be seen from the measured current-voltage (IV) characteristic which has a low voltage tail at currents close to $I_c$ and a sharp
voltage jump at large currents (see Fig. 4 in Ref. \cite{Probst}) which resembles the IV curves of wide superconducting bridges with $w\gg \lambda^2/d$
(see Fig. 2 in \cite{Zolochevskii}). Width of the used bridge ($w=4.5 \mu m$) satisfies the condition
$w \gg \xi \tau_{in}/\tau_{GL}$ (where we used $\tau_{in}\sim 1.7 \cdot 10^{-11} s$ at $T=0.9T_c$ \cite{Doettinger} and
$\xi(T=0.9 T_c)\sim 5 nm$) which means that the main contribution to the time delay comes from the creation process of a quasi-phase slip line.
It is clear that the last time depends not only on the intrinsic parameters of the material ($\tau_{in}$) but also on the width of the bridge.

And finally, we discuss the application of our results to the problem of the time delay in the appearance of the voltage response after
the absorption of a single photon in a superconducting single photon detector. The absorbed photon creates a hot spot
(region with suppressed $|\Delta|$) in the superconducting
film and the superconducting current density is redistributed in order to avoid this place \cite{SSPD_review}.
The current redistribution occurs on the time scale $\sim \tau_J$ which is much smaller than any inelastic relaxation times.
Therefore, it is safely to assume that the current distribution changes suddenly. Due to current crowding effect the distribution of the
current density will be nonuniform with the local maxima near the hot spot \cite{Zotova}. If the maximal current density exceeds
$j_{dep}$ the superconducting state becomes unstable and a finite voltage drop appears in the film after some time delay.
In Ref. \cite{Zhang} authors used the results of the uniform model \cite{Tinkham_review} to analyze their experimental results on the time delay.
Our calculations show that the spatially nonuniform current distribution does not change {\it qualitatively} the dependence of $t_d$ on
the current but it reduces $t_d$ considerably. Therefore usage of simple expressions following from the Tinkham's model may considerably underestimate
the actual value of $\tau_{in}$. Besides one should take into account that when the maximal current density
near the hot spot approaches $j_{c1}$ the time delay drops fast and it becomes about $\tau_{GL}$ (but in reality this time will be larger and
it is mainly determined by the creation time of the hot spot).

\section{Conclusion}

In our work we studied the dynamic response of 1D and 2D superconducting bridges after the abrupt switching on of the supercritical current.
We presented calculations near the critical temperature of the superconductor in two limits: i) non-thermal limit, when the energy relaxation
time of electrons due to electron-phonon interaction is the shortest one and the phonons are assumed to be in equilibrium, and
ii) quasi-equilibrium limit when the energy relaxation time of electrons is determined by the escape time of the nonequilibrium phonons to the substrate
and one can use the local temperature approach. We find that in both limits the fastest decay of the superconducitng order parameter
occurs near the ends of the homogenous bridge or, for a weakly inhomogenous bridge, in defect places where the local critical current is smallest.
We find that the time decay of $|\Delta|$ is smaller than in the model with spatially uniform suppression of $|\Delta|$ due to the diffusion
of nonequilibrium ('cooled') quasiparticles from the region where $|\Delta|$ decays faster. Smearing of the density of states and the spectral
functions at energies close to $|\Delta|$  (arising from finite supervelocity) is another factor which leads to a decrease of $t_d$.
Time delay does not depend on the temperature in the quasi-equilibrium limit (at currents slightly exceeding $I_c$) which is in strict
contrast with the non-thermal model which predicts a strong temperature dependence of $t_d$ taken at the same values $I/I_c(T)$.

Dynamic response of the 2D bridge with nonuniform current distribution resembles the response of the bridge with uniformly distributed current
but it has also some qualitative differences. The superconducting state in a 2D bridge is destroyed by the appearance of vortices and the subsequent
nucleation of the quasi-phase slip line across the bridge. For not very wide films $w/\xi \lesssim \tau_{in}/\tau_{GL}$ the nucleation time
of the first vortex and nucleation time of the quasi-phase slip line are close to each other at currents not far from $I_c$. By varying the
weak applied magnetic field one may change the level of nonuniformity in the current distribution and tune the time delay.

A voltage response of the superconducting bridge on an alternating current pulse (sinusoid like) is predicted if the duration of the pulse is larger
than the time delay after the abrupt switching of the current with magnitude equal to the amplitude of the ac current pulse. We also find that
the time-averaged voltage is not zero despite the zero time-averaged current and its sign changes when the phase of the ac current is changed by $\pi$.

\begin{acknowledgments}

The work was supported partially by the Russian Foundation for Basic Research (project 12-02-00509), by the Ministry of education and science of
the Russian Federation (project 8686) and by the European Science Foundation (ESF) within the framework of the activity entitled
'Exploring the Physics of Small Devices (EPSD)' (project 4327).

\end{acknowledgments}

\appendix

\section{Derivation of the heat conductance equation}

Here we derive the heat conductance equation which governs the dynamics of the local temperature of the quasiparticles in the quasi-equilibrium limit
when the electric field $E$ is small (or $E$ is finite only during a short time interval $\lesssim \tau_{GL}$) and one may neglect the Joule heating.
First of all from Eq. (4) it follows that

\begin{equation}\label{A1}
\frac{\partial N_1}{\partial t}+\frac{\partial|\Delta|}{\partial t}\frac{\partial R_2}{\partial \epsilon}-\frac{4Dq_s}{\hbar}
\frac{\partial q_s}{\partial t} \frac{\partial R_2N_2}{\partial \epsilon}=0.
\end{equation}

In the next step we multiply Eq. (\ref{A1}) by $f_L$, add it to the equation for $f_L$ (Eq. (2)) and than multiply the final equation by $N_0\epsilon$ and
integrate over the energy. As a result we obtain (with the help of the self-consistency equation $\Delta/\nu=1/2\int R_2f_Ld\epsilon$, where
$\nu$ is a coupling constant) the equation for the energy balance (per unit volume)
\begin{widetext}
\begin{equation}\label{A2}
\frac{\partial}{\partial t}\left[2N_0\int_{-\infty}^{\infty}\epsilon N_1f_Ld\epsilon-|\Delta|^2/\nu \right]=2N_0D \nabla\left(\int_{-\infty}^{\infty}
\epsilon(N_1^2-R_2^2)\nabla f_L d\epsilon \right) -\frac{2N_0}{\tau_{in}}\int_{-\infty}^{\infty}\epsilon N_1(f_L-f_L^0) d\epsilon.
\end{equation}
\end{widetext}
In Eq. (\ref{A2}) the term in brackets on the LHS corresponds to the energy of the electrons per unit volume $\mathcal{E}=F+TS$
(S is the entropy and F is the free energy per unit volume). This equation (here with omitted terms which incorporate effects of Joule dissipation and
coupling between transverse $f_T$ and longitudinal $f_L$ parts of $f(\epsilon)$) was originally derived in \cite{Schmid2} (see Eq. (C.4) there).
We seek for the solution of this equation in the form
\begin{equation} \label{A3}
f_L(\epsilon)=tanh(\epsilon/2k_BT_{loc}),
\end{equation}
and insert it in Eq. (\ref{A2}). After integration over energy we find the equation for the temperature of quasiparticles
\begin{widetext}
\begin{equation}\label{A4}
\frac{\partial}{\partial t}\left(\frac{N_0\pi^2k_B^2T_{loc}^2}{3}-N_0|\Delta|^2\frac{T_{loc}}{T_c}\right)=\frac{2N_0D\pi^2k_B^2}{3}\nabla \left(T_{loc}
\nabla T_{loc}\right)-\frac{N_0\pi^2k_B^2}{3}\frac{T_{loc}^2-T^2}{\tau_{in}}.
\end{equation}
\end{widetext}
In the derivation of Eq. (\ref{A4}) we used the Ginzburg-Landau expression for the free energy $F$, the expression for the entropy $S$
of the superconductor
near $T_c$ ($S=2\pi^2N(0)k_B^2T/2-N(0)|\Delta|^2/T_c$) and assume that $\Delta$ satisfies the time-dependent Ginzburg-Landau equation (Eq. (3)).
We also neglect terms $\sim |\Delta|^4$ in $\mathcal{E}$ which leads to $|\Delta|^4/\partial t \sim \Delta_{eq}^4/\tau_{GL}\sim (1-T/T_c)^3$
in the RHS of Eq. (\ref{A4}) and which is small near $T_c$ (note that it has the same smallness as the neglected Joule dissipation
$jE \sim \sigma_n j_{dep}^2\sim(1-T/T_c)^3$). In the framework of Eq. (\ref{A4}) the cooling of quasiparticles due to
a decreasing $|\Delta|$ has a simple physical origin - because of energy conservation the temperature of quasiparticles should go down
to compensate the energy increase due to the suppression of $|\Delta|$.

When the deviation from the equilibrium temperature is small $|T_{loc}-T|=|\delta T_{loc}|\ll T$ one may linearize Eq. (\ref{A4})
and we arrive at Eq. (6).


\end{document}